\documentclass[prl,twocolumn,showpacs,preprintnumbers,amsmath,amssymb]{revtex4}

\usepackage{epsfig}
\usepackage{graphicx}
\usepackage{dcolumn}
\usepackage{bm}

\newcommand{\EQ}{\begin{equation}}
\newcommand{\EN}{\end{equation}}
\newcommand{\ea}{\end{eqnarray}}
\newcommand{\ba}{\begin{eqnarray}}

\newcommand{\bear}{\begin{eqnarray}}
\newcommand{\ear}{\end{eqnarray}}

\begin{document}

\title{Neutron scattering in the hole-doped cuprate superconductors}
\author{J. M. P. Carmelo} 
\affiliation{GCEP-Centre of Physics, University of Minho, Campus Gualtar, P-4710-057 Braga, Portugal}
\date{3 May 2010}


\begin{abstract}
The occurrence of a neutron resonance energy is a common feature of unconventional superconductors.
In turn, the low-temperature incommensurate sharp peaks observed in the inelastic neutron scattering of 
La$_{2-x}$Sr$_x$CuO$_4$ (LSCO) correspond to four rods symmetrically distributed around $[\pi,\pi]$. 
Here it is shown that within the virtual-electron pair quantum liquid recently introduced the neutron 
resonance energy and the LSCO low-temperature incommensurate sharp peaks are generated by 
simple and closely related spinon processes. Our results indicate that in LSCO the neutron resonance 
energy either does not occur or corresponds to a lower energy $\approx 17$ meV.
\end{abstract}

\maketitle

Evidence is provided in Ref. \cite{Yu-09} that the occurrence of a neutron resonance energy
in different classes of unconventional superconductors follows from superconductivity 
being mediated by magnetic fluctuations. The spin-wave spectrum of the parent compound 
La$_2$CuO$_4$ (LCO) \cite{LCO-neutr-scatt} is quantitatively described by the square-lattice
quantum liquid (SLQL) of Ref. \cite{companion2}. Such a liquid refers to the Hubbard model on
the square lattice with effective transfer integral $t$ and on-site repulsion $U$,
defined in a subspace that contains its one- and two-electron excitations.
Within the SLQL the spinon processes that generate the spin-wave spectrum are very
simple, yet they lead to the same spectrum as Ref. \cite{LCO-Hubbard-NuMi}, where it is
derived by summation of an infinite set of ladder diagrams. The virtual-electron pair quantum liquid (VEPQL) 
recently introduced in Ref. \cite{cuprates0} refers to the SLQL perturbed by small
three-dimensional uniaxial anisotropy. Within it, a long-range superconducting order
emerges, mediated by magnetic fluctuations.
The description of both the SLQL and VEPQL in terms of charge $c$ fermions,
spin-singlet two-spinon $s1$ fermions, and independent spinons \cite{companion2,cuprates0}
profits from the extended global $SO(3)\times SO(3)\times U(1)=[SO(4)\times U(1)]/Z_2$ symmetry
found recently in Ref. \cite{bipartite} for the Hubbard model on any bipartite lattice.
The new found $U(1)$ symmetry is that broken at the critical temperature $T_c$.
In Refs. \cite{cuprates0,cuprates} the VEPQL is shown to successfully describe the
universal unusual properties of the hole doped cuprates, including the 
normal-state linear-$T$ resistivity \cite{cuprates}. 

In this Letter strong evidence is provided that combination of the simple $s1$ fermion processes studied
in Ref. \cite{companion2} for hole concentration $x=0$ with the contraction and deformation 
of the $s1$ boundary line upon increasing $x$ leads for approximately $x\in (x_c,x_{c1})$ to sharp peaks in 
the spin-triplet spectral weight distribution. They correspond to four rods symmetrical 
distributed about $\vec{\pi}=[\pi,\pi]$, alike those observed in the inelastic neutron scattering of La$_{2-x}$Sr$_x$CuO$_4$ (LSCO) 
\cite{Cuprates-insulating-phase,neutron-Yamada}. Here $x_c\approx 0.05$ and $x_{c1}\approx 1/8=0.125$.
Importantly, a second type of closely related processes whose spectrum we evaluate
for approximately $x\in (x_c,x_{op})$ where $x_{op}\approx 0.16$ leads to a sharp spectral feature centered at $\vec{\pi}$. 
Within the VEPQL scheme it is identified with the neutron resonance 
energy observed in the four representative systems other than LSCO considered in the
studies of Ref. \cite{cuprates0}: YBa$_2$Cu$_3$O$_{6+\delta}$ (YBCO 123), 
Bi$_{2}$Sr$_2$CaCu$_2$O$_{8+\delta}$ (Bi 2212), HgBa$_2$CuO$_{4+\delta}$ (Hg 1201), 
and Tl$_2$Ba$_2$CuO$_{6+\delta}$ (Tl 2201). 
(Due to disorder effects \cite{rotation}, for $x<x_c$ the momentum-space 
position of the peaks predicted here for LSCO is rotated by an angle $\pi/4$.)

For the $x=0$ and spin-density $m=0$ absolute ground state the $s1$ band coincides with an 
antiferromagnetic reduced Brillouin zone such that $\vert q_{x_1}\vert+\vert q_{x_2}\vert\leq\pi$.
Its $s1$ boundary-line momenta ${\vec{q}}_{Bs1}$ have Cartesian components obeying 
the equations $q_{Bs1x_1}\pm q_{Bs1x_2}=\pm, \mp\pi$ \cite{companion2}.
In turn, the following equations derived in Ref. \cite{companion2} for small values of $x$ and $U/4t\geq u_0$,
which due to the $s1$ boundary line contraction upon increasing $x$ are obeyed by the 
Cartesian components of the momenta belonging to that line and pointing in and near the nodal directions, remain a 
good approximation for $x\in (x_c,x_{c1})$,
\begin{equation}
q_{Bs1x_1} \pm q_{Bs1x_2} = \pm ,\mp(1 - x)\pi 
\, , \hspace{0.15cm} U/4t\geq u_0\approx 1.302 \, .
\label{qx_1-qx_2}
\end{equation}
(Unlike for one-electron excitations, for the spin excitations considered here
the $s1$ band momenta are independent of the doublicity $d=\pm 1$ \cite{companion2,cuprates0}.)
The $s1$ band is full for the initial $m=0$ ground state and displays two holes
for the spin excitations. The corresponding VEPQL general spin spectrum 
derived in Ref. \cite{cuprates0} reads,
\begin{eqnarray} 
\delta E_{spin} & = & -\epsilon_{s1} ({\vec{q}}\,') - \epsilon_{s1} ({\vec{q}}\,'') \, ; \hspace{0.35cm} 
\delta \vec{P} = [\delta {\vec{q}}_{c}^{\,0}-{\vec{q}}\,'-{\vec{q}}\,''] \, ,
\nonumber \\
\delta {\vec{q}}_{c}^{\,0} & = & 
\pm\left[\begin{array}{c}
\pi (1-x) \\ 
\pm\pi (1-x)
\end{array} \right]  
\, , \hspace{0.15cm} x=0 \, , \hspace{0.05cm}
x > x_c \, .
\label{DE-spin}
\end{eqnarray}
In the energy spectrum $\epsilon_{s1} ({\vec{q}})$ denotes the $s1$ fermion energy dispersion \cite{companion2,cuprates0}.
Our results refer to momenta ${\vec{q}}\,'\approx {\vec{q}}_{Bs1}'$ and
${\vec{q}}\,'\approx {\vec{q}}_{Bs1}''$ at or near the $s1$
boundary line for which according to the results of Ref. 
\cite{cuprates0}, $-\epsilon_{s1} ({\vec{q}}\,') - \epsilon_{s1} ({\vec{q}}\,'')\approx 
\vert\Delta\vert\vert\cos 2\phi' + \cos 2\phi''\vert +2\vert\Omega\vert\vert\sin 2\phi\vert$. Here the angles
$\phi'$, $\phi''$, and $\phi$ are associated with the directions of ${\vec{q}}\,'$,
${\vec{q}}\,''$, and ${\vec{q}} =[1/2]({\vec{q}}\,'-{\vec{q}}\,'')$, respectively.
Moreover, $\vert\Delta\vert= g_1\,\Delta_0$ and $\vert\Omega\vert =
\gamma_d\, g\, \Delta_0/(1-[x/x_*][4g/\gamma_c])$ 
are the spinon and superconducting VEP pairing energy per electron, respectively, \cite{cuprates0}.
The amplitudes $g_1 = \vert\langle e^{i\theta_{j,1}}\rangle\vert$ and
$g= \vert\langle e^{i\theta_j}\rangle\vert=\vert\langle e^{i[\theta_{j,0}+\theta_{j,1}]}\rangle\vert$
in such energy expressions are averages over the whole system of phase factors fully controlled by the quantum
fluctuations of the VEP phases $\theta_{j,1}$ and $\theta_{j} = \theta_{j,0}+\theta_{j,1}$,
respectively, $\gamma_c=(1-x_c/x_*)$, and $\gamma_d$ 
such that $\gamma_c<\gamma_d<1$ is a suppression coefficient. 
The latter coefficient accounts for the effects of intrinsic disorder or superfluid-density
anisotropy, which for $x\in (x_c,x_*)$ are small  \cite{cuprates0}. 
The parameters appropriate to the representative systems are $U/4t\approx 1.525$,
$t\approx 295$ meV, $x_c\approx 0.05$, and $x_*\approx 0.27$, so that
$\Delta_0\approx 42$ meV for LSCO and $\Delta_0\approx 84$ meV 
for the four systems other than LSCO \cite{cuprates0}. The VEPQL
critical temperature $T_c$ above which there is no long-range
superconducting order and the pseudogap temperature $T^*$ above which
there is no short-range spin order read
$T_c = \gamma_d\,{\breve{g}}\,\Delta_0/2k_B$ and
$T^* \leq {\breve{g}}_1[\Delta_0/k_B]$, respectively. 
Here ${\breve{g}}=g\vert_{T=0}\approx [(x-x_c)/(x_*-x_c)]\,{\breve{g}}_1$ 
and ${\breve{g}}_1=g_1\vert_{T=0}\approx (1-x/x_*)$ for $x\in (x_c,x_*)$.

Within the present status of both the SLQL and
VEPQL schemes one cannot calculate explicitly matrix elements of the two-electron 
spin-triplet operator between energy eigenstates and corresponding spectral-weight distributions.
The $x=0$ and $m=0$ results of Ref. \cite{companion2} on the spin-wave 
spectrum of the parent compound LCO \cite{LCO-neutr-scatt} profit from combination
of the SLQL with the complementary
method of Ref. \cite{LCO-Hubbard-NuMi}. They reveal that  
the microscopic mechanisms that generate the coherent spectral-weight spin-wave energy spectrum are 
in terms of two-spinon $s1$ fermion processes very simple. Indeed the two-spinon $s1$ fermion description
renders a complex many-electron problem involving summation of an infinite set of ladder diagrams
\cite{LCO-Hubbard-NuMi}
into a non-interacting two-$s1$-fermion-hole spectrum, described by simple analytical expressions. That
is consistent with the momentum values of the $c$ and $s1$ fermions being good quantum 
numbers for the SLQL. At $x=0$ and $m=0$ 
the spin spectrum (\ref{DE-spin}) refers to an effectively two-$s1$-band-hole non-interacting problem. The $c$ 
fermion band remains full in the excited states and thus there is no $c$ Fermi line. Only two holes emerge in the $s1$ band. 
The exclusion principle, phase-space restrictions, and momentum and energy conservation then drastically limit 
the number of available momentum occupancy configurations of the final excited states. In turn, the one-electron problem studied in 
Ref. \cite{cuprates} is more complex. For $x>x_c$ its physics is determined by 
low-energy and small-momentum inelastic $c$ - $s1$ fermion scattering near the isotropic 
$c$ Fermi line and anisotropic $s1$ boundary line, respectively. 

At $x=0$ and $m=0$ the sharp features of the spin spectral-weight distribution
are generated by processes corresponding to specific values of the two momenta 
${\vec{q}}\,'$ and ${\vec{q}}\,''$ of the general spin spectrum 
(\ref{DE-spin}). Such sharp features refer to coherent $\delta$-peak like spectral weight. Out of those, the 
sharpest spin peaks contain most coherent spin-wave spectral weight. They 
are generated by processes under which two $s1$ fermion holes emerge at momenta 
near the intersection of the lines $[q_{x_1},\pm\pi/2]$ or $[\pm\pi/2,q_{x_2}]$ with the $s1$
boundary line. At $x=0$ this refers to $s1$ band momenta belonging to that
line and pointing in the nodal directions. (Within the spectra of Eqs. (140) and (142) of Ref. \cite{companion2},
they are the $s1$ band momenta of components ${\vec{q}}\,'=[-\pi/2,\pi/2]$ and ${\vec{q}}\,''=-{\vec{q}}\,'=[\pi/2,-\pi/2]$ 
referring to $k_x=k_y=\pi$ and $k_y=\pi$, respectively.) While the momentum parts corresponding
to the motion of the center of mass of the broken spinon pair vanish, the relative momenta 
belong to the $s1$ boundary line and point
in the nodal directions. The resulting sharpest spin spectral feature is centered at the 
momentum $\vec{\pi}=[\pi,\pi]$. Most of the $x=0$ spin coherent spectral weight is concentrated at and near 
that sharpest spin peak. (See Fig. 3 (B) of Ref. \cite{LCO-neutr-scatt} for LCO.)

We now provide evidence that also for $x\in (x_c,x_{c1})$ and 
$m=0$ the $c$ fermion and two-spinon $s1$ fermion description of Refs. \cite{companion2,cuprates0}
renders a very complex many-electron spin problem into a much simpler one. 
(Some of our results remain valid for at least $x\in (x_c,x_{op})$.)
Indeed, the $c$ and $s1$ fermion momentum values are close to good 
quantum numbers for the VEPQL used in our studies \cite{cuprates0}.
Our results are valid for $U/4t\geq u_0\approx 1.3$. 
The lack of long-range antiferromagnetic order in the initial $x\neq 0$ and $m=0$
ground state \cite{companion2,cuprates0} implies that now the sharpest spectral features do not
refer to coherent spin weight and thus are not $\delta$-function like. Our main conjecture is
that for hole concentrations approximately in the range $x\in (x_c,x_{c1})$ such sharpest spin 
features are generated as well by processes under which: (i) The two $s1$ fermion holes 
in the general spin spectrum of Eq. (\ref{DE-spin}) emerge at momenta ${\vec{q}}\,'$ and ${\vec{q}}\,''$ 
in the intersection of the lines $[q_{x_1},\pm\pi/2]$ or $[\pm\pi/2,q_{x_2}]$ with the $s1$
boundary line; (ii) The momentum part ${\vec{q}}=[1/2]({\vec{q}}\,'-{\vec{q}}\,'')$
of such momenta corresponding to the spinon relative motion in 
the broken pair is a $s1$ boundary-line nodal momentum. It follows that both the momenta
${\vec{q}}\,'$ and ${\vec{q}}\,''$ and the corresponding relative momentum
${\vec{q}}=[1/2]({\vec{q}}\,'-{\vec{q}}\,'')$ must belong to the $s1$ boundary line. Both the conditions (i) and (ii) are
met at $x=0$. For $x\in (x_c,x_{c1})$ such a conjecture is confirmed below to lead to results 
fully consistent with the low-temperature neutron
inelastic scattering of the five representative hole-doped cuprate superconductors. 
The incommensurability effects bring about a finite momentum part 
$\delta {\vec{q}}=({\vec{q}}\,'+{\vec{q}}\,'')$ referring to the 
motion of the center of mass of the broken spinon pair.

Following our above discussion, we consider that alike for 
$x=0$ and $m=0$, for $x\in (x_c,x_{c1})$ and $m=0$ the sharpest spin 
spectral peaks refer to processes meeting the above conditions (i) and (ii). 
From use of Eq. (\ref{qx_1-qx_2}), one finds that 
the condition (i) that the momenta ${\vec{q}}\,'$ and ${\vec{q}}\,''$ 
are in the intersection of the lines $[q_{x_1},\pm\pi/2]$ or $[\pm\pi/2,q_{x_2}]$ with the $s1$
boundary line is met by $s1$ boundary-line momenta ${\vec{q}}\,'$
and ${\vec{q}}\,''$ of the general form,
\begin{equation}
{\vec{q}}^{\,N,a,l,l'}_{Bs1} =
l\left[\begin{array}{c}
\pi/2 \\ 
l'(\pi/2 - x\,\pi)
\end{array} \right]  
; {\vec{q}}^{\,N,b,l,l'}_{Bs1} =
l\left[\begin{array}{c}
l'(\pi/2 -  x\,\pi) \\ 
\pi/2 
\end{array} \right] 
\label{qBx_1-qBx_2}
\end{equation}
where $l, l' =\pm 1$. Furthermore, the condition (ii) that the 
corresponding relative momentum ${\vec{q}}=[1/2]({\vec{q}}\,'-{\vec{q}}\,'')$
is a $s1$ boundary-line nodal momentum is met by
the momentum pairs ${\vec{q}}\,'= {\vec{q}}^{\,N,a,l,l'}_{Bs1}$ and ${\vec{q}}\,'={\vec{q}}^{\,N,b,-l\,l',l'}_{Bs1}$.
Indeed, such momenta can be written as,
\begin{equation}
{\vec{q}}^{\,N,a,l,l'}_{Bs1} =
{\vec{q}}^{\,N}_{Bs1} + {1\over 2}
\delta{\vec{q}}^{\,N,l,l'}_{Bs1}  ; 
{\vec{q}}^{\,N,b,-l\,l',l'}_{Bs1} =
- {\vec{q}}^{\,N}_{Bs1} + {1\over 2}
\delta{\vec{q}}^{\,N,l,l'}_{Bs1} 
\label{qBx_1-qBx_2-qN-dq}
\end{equation}
Here the relative momentum,
\begin{equation}
{\vec{q}}^{\,N}_{Bs1} =
l\left[\begin{array}{c}
{\pi\over 2} (1-x) \\ 
l' {\pi\over 2} (1-x)
\end{array} \right]  
\, ; \hspace{0.35cm} 
l, l' =\pm 1 \, ,
\label{qBqN-dq}
\end{equation}
is indeed a $s1$ boundary-line momentum pointing in nodal directions. It
corresponds to the relative motion of the spinons in the broken pair.
In turn, the momentum,
\begin{equation}
\delta{\vec{q}}^{\,N,l,l'}_{Bs1} \equiv 
{\vec{q}}^{\,N,a,l,l'}_{Bs1} + {\vec{q}}^{\,N,b,-l\,l',l'}_{Bs1}
= l\,l'\,x\pi \left[\begin{array}{c}
1 \\ 
-1
\end{array} \right] \, ,
\label{delta-q-ll-Bs1}
\end{equation} 
corresponds to the motion of the center of mass of such a pair. 
The momentum $\delta{\vec{q}}^{\,N,l,l'}_{Bs1}$ vanishes at $x=0$,
consistently with it vanishing for the processes leading to the $x=0$ sharpest
spin spectral features. 

There are two types of closely related $s1$ fermion processes for which ${\vec{q}}\,'= {\vec{q}}^{\,N,a,l,l'}_{Bs1}$ 
and ${\vec{q}}\,'={\vec{q}}^{\,N,b,-l\,l',l'}_{Bs1}$ in the general spin spectrum (\ref{DE-spin}).
For the first process type $\delta {\vec{q}}_{c}^{\,0}=\pm (1-x)[\pi,\pi]$ 
in that spin spectrum. We straightforwardly find the following
value for the momentum $\delta \vec{P}$ of Eq. (\ref{DE-spin})
valid for hole concentrations approximately in the range $x\in (x_c,x_{c1})$,
\begin{eqnarray}
\delta \vec{P} & = & \delta {\vec{q}}_{c}^{\,0} -
{\vec{q}}^{\,N,a,l,l'}_{Bs1} - {\vec{q}}^{\,N,b,-l\,l',l'}_{Bs1} =
{\vec{q}}_{c}^{\,0} - \delta{\vec{q}}^{\,N,l,l'}_{Bs1} 
\nonumber \\ 
& = & \pm\left[\begin{array}{c}
\pi (1-x) \\
\pi (1-x)
\end{array} \right] -
 l\,l'\,x\pi \left[\begin{array}{c}
1 \\ 
-1
\end{array} \right] 
\nonumber \\
& = & \pm\left[\begin{array}{c}
\pi  \\
\pi (1- 2x)
\end{array} \right] \, , \hspace{0.15cm} 
l\,l' =\mp 1 \, ,
\nonumber \\
& = & \pm\left[\begin{array}{c}
\pi (1- 2x) \\
\pi 
\end{array} \right] \, , \hspace{0.15cm} 
l\,l' =\pm 1 \, .
\label{K-qBx_1-qBx_2}
\end{eqnarray}
Owing to the lattice discrete translational symmetry, corresponding processes originated by 
adding to those a suitable reciprocal-lattice vector ${\vec{G}}^{\mp}=\mp [2\pi,2\pi]$ give rise to 
the excitation momenta,
\begin{eqnarray}
\delta \vec{P} & = & 
\pm\left[\begin{array}{c}
\pi (1-x) \\
\pi (1-x)
\end{array} \right] -
l\,l'\,x\pi \left[\begin{array}{c}
 1 \\ 
-1
\end{array} \right] 
\mp\left[\begin{array}{c}
2\pi \\
2\pi 
\end{array} \right]
\nonumber \\
& = & \mp\left[\begin{array}{c}
\pi  \\
\pi (1+ 2x)
\end{array} \right] \, , \hspace{0.15cm} 
l\,l' =\mp 1 \, ,
\nonumber \\
& = & \mp\left[\begin{array}{c}
\pi (1+ 2x) \\
\pi 
\end{array} \right] \, , \hspace{0.15cm} 
l\,l' =\pm 1 \, .
\label{K-G-qBx_1-qBx_2}
\end{eqnarray}
As discussed below, this first type of processes correspond to momentum
locations identical to those of the sharp 
low-temperature incommensurate peaks in the inelastic neutron scattering of LSCO. 

The corresponding incommensurability is defined as $\delta_{inc} \equiv [1/2\pi]\vert\vec{\pi}-\delta \vec{P}\vert$.
Here $\delta \vec{P}$ stands for any of the excitation momenta of Eqs. (\ref{K-qBx_1-qBx_2}) and
(\ref{K-G-qBx_1-qBx_2}). We now use the $s1$ boundary line nodal momentum $q^N_{Bs1}$ expression 
provided in Ref. \cite{cuprates0} to derive the following corresponding expression of the incommensurability valid for 
the extended hole concentration range $x\in (x_c,x_*)$,
\begin{eqnarray}
\delta_{inc} \equiv {1\over 2\pi}\vert\vec{\pi}-\delta \vec{P}\vert
& = & x \, , \hspace{0.35cm} x \in (x_c,x_{c1}) \, ,
\nonumber \\
& = & x_{c1} \, , \hspace{0.35cm} x \in (x_{c1},x_*) \, .
\label{delta-x}
\end{eqnarray}
Hence, for the present hole concentration range $x\in (x_c,x_{c1})$ and for instance for the quadrant for which the components
the momenta $\delta \vec{P}$ are such that $\delta P_{x_1},\delta P_{x_2}\geq 0$, the
sharpest peaks of the spin spectral weight distribution correspond within our above generalization of
the $x=0$ processes found in Ref. \cite{companion2} to four 2D
rods at $\delta \vec{P}=[\pi\pm 2\pi\delta_{inc},\pi]$ and $\delta \vec{P}=[\pi ,\pi\pm 2\pi\delta_{inc}]$.
Those are symmetrical distributed about $\vec{\pi}=[\pi,\pi]$. 

A second type of closely related processes that also obey the above conditions (i) and
(ii) refer to the choices $\delta {\vec{q}}_{c}^{\,0}= \pm (1-x)[\pi,-\pi]$, 
${\vec{q}}\,'= {\vec{q}}^{\,N,a,l,l'}_{Bs1}$, and ${\vec{q}}\,'={\vec{q}}^{\,N,b,-l\,l',l'}_{Bs1}$
in the general spin spectrum (\ref{DE-spin}) for $l\,l'=\mp 1$,
\begin{equation}
\delta \vec{P} =
\pm\left[\begin{array}{c}
\pi (1-x) \\
-\pi (1-x)
\end{array} \right] \pm x\pi \left[\begin{array}{c}
1 \\ 
-1
\end{array} \right] = \pm\left[\begin{array}{c}
\pi  \\
-\pi 
\end{array} \right] \, .
\label{K-qBx_1-qBx_2-pipi}
\end{equation}
This spectrum is expected to remain a good approximation at least for $x\in (x_c,x_{op})$.

The energy spectra corresponding to the momentum spectra
of Eqs. (\ref{K-qBx_1-qBx_2}) and (\ref{K-G-qBx_1-qBx_2}) and 
spectrum of Eq. (\ref{K-qBx_1-qBx_2-pipi}) have the same
form. From the use of the general spin spectrum of Eq. (\ref{DE-spin})
one finds for zero temperature,
\begin{eqnarray}
\delta E_{spin} & = & -\epsilon_{s1} ({\vec{q}}^{\,N,a,l,l'}_{Bs1}) - 
\epsilon_{s1} ({\vec{q}}^{\,N,b,-l\,l',l'}_{Bs1}) 
\nonumber \\
& \approx & 
\vert\Delta\vert\vert\cos 2\phi' + \cos 2\phi''\vert +2\vert\Omega\vert\vert\sin 2\phi\vert
\nonumber \\
& = & 2\vert\Omega\vert \, .
\label{E-spin-N}
\end{eqnarray}
Here the angles $\phi'$ and $\phi''$ and the angle $\phi$ are associated with the directions pointed
by the momenta ${\vec{q}}\,'= {\vec{q}}^{\,N,a,l,l'}_{Bs1}$ and ${\vec{q}}\,''={\vec{q}}^{\,N,b,-l\,l',l'}_{Bs1}$
of Eq. (\ref{qBx_1-qBx_2-qN-dq}) and the relative momentum ${\vec{q}}={\vec{q}}^{\,N}_{Bs1}$ of Eq. (\ref{qBqN-dq}), 
respectively. The surprisingly simple energy spectrum $\delta E_{spin} \approx 2\vert\Omega\vert$
associated with both the excitation momenta of Eqs. (\ref{K-qBx_1-qBx_2}) and (\ref{K-G-qBx_1-qBx_2}) and 
Eq. (\ref{K-qBx_1-qBx_2-pipi}) follows from $\cos 2\phi' = - \cos 2\phi''$ and $\vert\sin 2\phi\vert=1$ for 
the $s1$ band momenta under consideration. Within the VEPQL scheme the second type of processes whose 
spectrum is given in Eqs. (\ref{K-qBx_1-qBx_2-pipi}) and (\ref{E-spin-N}) generate the neutron resonance 
energy observed in YBCO 123, Bi 2212, Hg 1201, and Tl 2201, as discussed below.

For temperatures below $T_c$ and a well-defined temperature-dependent range 
the spectrum (\ref{E-spin-N}) has the same form except that the amplitude $g$
and thus the pairing energy $2\vert\Omega\vert$ decrease 
upon increasing $T$ and vanish for temperatures above $T_c$. Hence for temperatures 
in the range $T\in (T_c,T^*)$ below the pseudogap temperature $T^*$ 
the spin energy (\ref{E-spin-N}) vanishes, $\delta E_{spin} \approx 0$.

The low-temperature incommensurate peaks in the inelastic neutron scattering of LSCO 
are for $x\in (x_c,x_{c1})\approx (0.05,0.125)$ observed precisely at the momentum values 
$\delta \vec{P}=[\pi\pm 2\pi\delta_{inc},\pi]$ and $\delta \vec{P}=[\pi ,\pi\pm 2\pi\delta_{inc}]$
\cite{Cuprates-insulating-phase,neutron-Yamada} of Eqs. (\ref{K-qBx_1-qBx_2}) and (\ref{K-G-qBx_1-qBx_2}).
The LSCO experimental dependence of $\delta_{inc}$ on $x$ is studied in Refs. \cite{Cuprates-insulating-phase,neutron-Yamada}. 
For instance the Sr-doping experimental points of Fig. 37 (a) of Ref. \cite{Cuprates-insulating-phase}
confirm that for LSCO the theoretical magnitudes $\delta_{inc} \approx x$ for 
$x\in (x_c,x_{c1})\approx (0.05,0.125)$ and $\delta_{inc} \approx x_{c1}=1/8=0.125$ for 
$x\in (x_{c1},x_*)\approx (0.125,0.27)$ given in Eq. (\ref{delta-x}) are indeed approximately valid. 

At zero temperature the excitation energy corresponding to the excitation 
momenta $\delta \vec{P}$ of Eqs. (\ref{K-qBx_1-qBx_2}) and (\ref{K-G-qBx_1-qBx_2})
reads $\delta E_{spin}\approx 2\vert\Omega\vert$, as given in Eq. (\ref{E-spin-N}). 
Upon increasing the temperature $T$, the phase-coherent
virtual-electron pairing energy $2\vert\Omega\vert$ decreases and vanishes as $T\rightarrow T_c$.
Therefore, for $x\in (x_c,x_{c1})$ and for the low-energy $\delta E_{spin}\approx 0$
regime associated with the spectrum (\ref{E-spin-N}) for temperatures $T\geq T_c$ 
at which $2\vert\Omega\vert=0$
the intensity of the incommensurate peaks as a function of $T$ has a sharp maximum
at $T\approx T_c$. It diminishes rapidly toward zero as $T$ is lowered, consistently
with the opening of the spin-spectrum nodal gap associated with the phase-coherent
virtual-electron pairing energy $2\vert\Omega\vert$. It refers to the spectrum
$\delta E_{spin} \approx  2\vert\Omega\vert$ of Eq. (\ref{E-spin-N}). 
This is again the behavior observed in LSCO \cite{Cuprates-insulating-phase}.

In YBCO 123 the width of the scattering makes it difficult to observe
any incommensurability. In that material and also in Bi 2212, Hg 1201,
and Tl 2201 a single sharp spin spectral feature centered at $\vec{\pi}=[\pi,\pi]$ is observed. 
Within the VEPQL it refers to the spectrum given in Eqs. (\ref{K-qBx_1-qBx_2-pipi}) and (\ref{E-spin-N})
generated by the above second type of excitations. Their excitation energy 
is at low temperature given by $\delta E_{spin} \approx  2\vert\Omega\vert\approx 2\vert\Omega\vert\vert_{T=0}$.
Its maximum magnitude $\delta E_{spin} \approx  2\vert\Omega\vert^{max}\approx 2\vert\Omega\vert\vert_{T=0}^{max}$ 
occurs at the optimal hole concentration $x_{op}=0.16$. The theoretical energy
$2\vert\Omega\vert^{max}\approx \gamma_d^{min}\,\Delta_0/2\approx 5k_B\,T^{max}_c$ 
calculated for the parameters appropriate to YBCO 123, Bi 2212, Hg 1201,
and Tl 2201 reads $2\vert\Omega\vert^{max}\approx 40$ meV for all of them.
This agrees quantitatively with the neutron resonance energy
$2\vert\Omega\vert^{max}$ observed in such materials, which is given
by $39-41$ meV \cite{ARPES-review}.  

We emphasize that the two types of closely related processes considered here
only differ in the sign $\pm $ of the factor $[\pi,\pm \pi]$ appearing
in the momentum $\delta {\vec{q}}_{c}^{\,0}= \pm (1-x)[\pi,\pm\pi]$
of the general spin spectrum (\ref{DE-spin}). The first type of
excitation is difficult to observe in the four systems other
than LSCO due to the scattering width. The neutron resonance energy
was not observed in LSCO.
If the second type of excitations with spectrum given 
in Eqs. (\ref{K-qBx_1-qBx_2-pipi}) and (\ref{E-spin-N}) occurred in LSCO
the theoretical energy calculated for the parameters appropriate to
it would be $2\vert\Omega\vert^{max} \approx 17$ meV.
One possibility is that due to matrix-elements issues specific to the LSCO
physics only the excitations with spectrum given in
Eqs. (\ref{K-qBx_1-qBx_2}), (\ref{K-G-qBx_1-qBx_2}), and (\ref{E-spin-N})
generate sharp spin spectral features. Whether a smaller neutron resonance energy
$2\vert\Omega\vert^{max} \approx 17$ meV occurs in LSCO is a problem
that deserves further experimental investigations. 

Hence we conclude that within the conjecture that the sharpest spin spectral features 
are generated for $x\in (x_c,x_{c1})$ by a type of two-spinon $s1$ fermion processes similar to those found
in Ref. \cite{companion2} for $x=0$ and $m=0$, the VEPQL scheme of Refs. \cite{cuprates0,cuprates}
leads to remarkable quantitative agreement with
the low-temperature incommensurate peaks in the inelastic neutron scattering of LSCO \cite{Cuprates-insulating-phase}.
Excellent quantitative agreement between our theoretical predictions and the neutron resonance energy observed in
YBCO 123, Bi 2212, Hg 1201, and Tl 2201 is also achieved. 
Importantly, our results reveal that the low-$T$ incommensurate sharp peaks and the neutron resonance energy
observed in LSCO and the remaining four representative systems, respectively, are generated by closely related 
processes. They are two faces of the same phenomenon.
Our results strongly suggest that the $c$ and $s1$ fermion description used in the studies of this Letter
renders a very complex spin-spectrum many-electron problem into a much simpler problem in terms of 
suitable two-$s1$-fermion-hole $s1$ band processes. This is consistent with the VEPQL momentum values
of the $c$ and $s1$ fermions being close to good quantum numbers \cite{cuprates0,cuprates}.

\begin{acknowledgments}
I thank M. A. N. Ara\'ujo, A. Damascelli, P. D. Sacramento, M. J. Sampaio, and K. Yamada for discussions 
and the support of the ESF Science Program INSTANS and grant PTDC/FIS/64926/2006.
\end{acknowledgments}

\end{document}